%% file: current.tex
\documentclass[article]{IEEEtran}
\usepackage[utf8]{inputenc}
\usepackage{graphicx} 
\usepackage{multirow}
\usepackage{multicol} 
\usepackage{booktabs}
\usepackage{amssymb}
\usepackage{pifont}

\usepackage{nopageno} 
\usepackage{caption}
\usepackage{subcaption}
\usepackage{algorithm}
\usepackage{algpseudocode}
\usepackage{bm}
\usepackage{amsmath}
\usepackage{mathtools}
\usepackage{commath}
\usepackage{nomencl}

\usepackage{tikz}
\usepackage{pgf}
\usepackage{pgfplots}
\usepackage{tikz-network}
\usepackage{authblk}

\usepackage{float}
\newfloat{algorithm}{t}{lop}

\graphicspath{{Figures/}}

\DeclareMathOperator*{\argmin}{argmin}
\DeclareMathSymbol{\ThetaCal}{\mathord}{operators}{"02}
\algnewcommand{\Initialize}[1]{%
  \State \textbf{Initialize}
  \Statex \hspace*{\algorithmicindent}\parbox[t]{.8\linewidth}{\raggedright #1}
}
\algnewcommand{\Solve}[1]{%
  \State \textbf{Solve}
  \Statex \hspace*{\algorithmicindent}\parbox[t]{.8\linewidth}{\raggedright #1}
}
\algnewcommand{\RequireR}[1]{%
  \State \textbf{Require}
  \Statex \hspace*{\algorithmicindent}\parbox[t]{.8\linewidth}{\raggedright #1}
}
\newcommand*{\argminl}{\argmin\limits}

\renewcommand{\footnoterule}{%
\kern -3pt
\hrule width 2in
\kern 2.6pt
}

\input{include}

\title{Unveiling Anomalous Edges and Nominal\\ Connectivity of Attributed Networks \thanks{\;\;The work in this paper was supported by the NSF grants 1901134, 171141, and 1500713.} \thanks{\;\;Emails: {\textit {\{polyz003, mavro016, ioann006, georgios\}@umn.edu}}
}}

\author[$\ast$,1]{Konstantinos D. Polyzos\thanks{$^\ast$Equal Contribution.}}
\author[$\ast$,2]{Costas Mavromatis}
\author[1]{Vassilis N. Ioannidis}
\author[1]{Georgios B. Giannakis}
\affil[1]{Department of Electrical and Computer Engineering, University of Minnesota, USA}
\affil[2]{Department of Computer Science, University of Minnesota, USA}

\begin{document}
\maketitle
\begin{abstract}
Uncovering anomalies in attributed networks has recently gained popularity due to its importance in unveiling outliers and flagging adversarial behavior in a gamut of data and network science applications including {the Internet of Things (IoT)}, finance, security, to list a few. The present work deals with uncovering anomalous edges in attributed graphs using two distinct formulations with complementary strengths, which can be easily distributed, and hence efficient.  The first relies on decomposing the graph data matrix into low rank plus sparse components to markedly improve performance. The second broadens the scope of the first by performing robust recovery of the unperturbed graph, which enhances the anomaly identification performance. The novel methods not only capture anomalous edges linking nodes of different communities, but also spurious connections between any two nodes with different features. Experiments conducted on real and synthetic data corroborate the effectiveness of both methods in the anomaly identification task. 

\end{abstract}


\thispagestyle{empty} 

\section{Introduction}

Unveiling rare or anomalous data is of paramount importance in numerous applications such as recommender systems, finance, security, health care, autonomous driving and insurance~\cite{Akoglu2014Graph-basedSurvey}.  Anomaly identification methods can spot  network intrusions and failures, spam reviews, credit card frauds, malware and rare disease outbreaks. 

The present work focuses on anomalies that appear in data adhering to a graph structure. Consider a graph that consists of nodes and edges, where each node is associated with an attributed vector. In an IoT network for example, the nodes represent mobile devices and the edges capture the communication among devices. The graph connectivity may be perturbed e.g. by an adversary who adds edges possibly at random. Examples of such anomalous edges include links among nodes belonging to different communities or among nodes with uncorrelated attributes. Leveraging the graph structure and the correlation of nodal features will enable identification of anomalous edges.


Anomalies can emerge in either static or dynamic graphs. Dynamic ones comprise sequences of graphs and the corresponding methods leverage temporal correlation too~\cite{Mardani2013DynamicRank, Ioannidis2018CoupledDetection,Baingana2016JointNetworks, KoutraTensorSplat}. The contemporary work in~\cite{ceci2020graph,ceci2020graphunc,ceci2018signal} considers erroneous signal on the graph and edges and insights from there can also be utilized to extend and complement the work in this paper.  The focus here will be on static graphs, but dynamic graphs are also included in our future research agenda. 

Most anomaly detection methods for static graphs pursue community-based approaches. 
They primarily rely on matrix factorization techniques that decompose the graph matrix into a low rank component and a residual matrix \cite{Chakrabarti2004AutoPart:Detection,Tong2011Non-negativeDetection}.
Matrix factorization is widely employed for dimensionality reduction \cite{Nikulin2012UnsupervisedRates, Ambai2011PAPERFactorization} and (graph) clustering \cite{Kuang2012SymmetricClustering}. The approach in \cite{Tong2011Non-negativeDetection} spots anomalous connections if they appear in the residual matrix, after the non-negative matrix factorization. 

Albeit interesting, the aforementioned methods do not account for attributes. On the other hand, approaches have been reported that incorporate nodal features to unveil anomalous nodes \cite{Gao2010ADistance, Muller2013RankingGraphs, Akoglu2010OddBall:Graphs, ioannidis2020graphsac,ioannidis2020defending,ioannidis2020efficient,ioannidis2020tensorgcn}. These are suitable to spot anomalous nodes within a certain community, but not anomalous edges across communities \cite{Gaot2010OnNetworks}. 

\textbf{Contribution.} The present work proposes two novel approaches that rely on low rank and sparse factorization, and judiciously account for nodal attributes to spot anomalous edges in a given perturbed graph. The second approach further recovers the ground-truth topology, which turns out to markedly enhance identification performance. Accounting for smoothness across nodal features also 
boosts identification performance. Different from existing approaches, the novel methods not only identify anomalous links between different communities, but also mendacious edges between any two nodes based on the nodal features. Finally, two optimization methods with complementary strengths are developed to unveil the anomalous edges and robustly recover the underlying nominal topology. Although the first method is computationally more efficient, the second unveils more accurately the perturbed edges. Complexity analysis of the developed methods is also provided.

\section{Unveiling anomalous edges}

Consider a graph that consists of $N$ nodes and $E$ edges connecting pairs of nodes. Connectivity is captured by the $N\times N$ adjacency matrix $\mathbf{A}$. The initial (a.k.a. nominal) graph is perturbed by inserting (anomalous) edges linking nodes with correlated or uncorrelated nodal features. Let $\tilde{\mathbf{A}}$ denote the perturbed adjacency, and  $\tilde{\mathcal{E}}$ the set of perturbed edges. The $N \times N$ Laplacian matrix is 

\begin{equation}
 \mathbf{L} := \text{diag}(\mathbf{A}\mathbf{1})-\mathbf{A}
\label{eq:Laplmatrix}
\end{equation}
where $\mathbf{1}$ is the all-ones vector and $\text{diag}(\cdot)$ is a diagonal matrix.

Each node $n$ also holds an $F\times 1$ feature vector $\mathbf{x}_n$, and all feature vectors are collected in the $N\times F$ feature matrix  $\mathbf{X}$. For example, in a network that consists of voters, attributes may correspond to each voter's age and state of residency.

\begin{figure}
\begin{subfigure}{.24\textwidth} 
\begin{tikzpicture}
\Vertex[size=.3,x=-.5,y=2.2,color=blue,label=Democrat, position=right]{0}
\Vertex[size=.3,x=-.5,y=1.8,color=red,label=Republican, position=right]{1}
\Vertex[size=.1,x=1.5,y=2.2,Pseudo=True]{2a}
\Vertex[size=.1,x=2,y=2.2,Pseudo=True, label=edge, position = right]{3a}
\Vertex[size=.1,x=1.5,y=1.8,Pseudo=True]{2}
\Vertex[size=.1,x=2,y=1.8,Pseudo=True, label=anomalous edge, position = right]{3}
\Edge(2a)(3a)
\Edge[color=red,style={dashed}](2)(3)

\Vertex[size=.3,x=-.2,color=blue,label=$\text{[25,CA]}$, position=above]{A}
\Vertex[size=.3,x=.5,y=1, color=blue, label=$\text{[30,CA]}$, position=above]{B}
\Vertex[size=.3,x=.5,y=-1, color=blue,label=$\text{[35,CA]}$, position=below]{C}
\Vertex[size=.3,x=1.2, color=blue,label=$\text{[40,NV]}$, position=below]{D}
\Vertex[size=.3,x=2.2, color=red,label=$\text{[45,NV]}$, position=below]{E}
\Vertex[size=.3,x=2,y=1, color=red,label=$\text{[35,UT]}$, position=above]{F}
\Vertex[size=.3,x=3,y=1, color=red,label=$\text{[20,ID]}$, position=above]{H}
\Vertex[size=.3,x=3,y=-1, color=red,label=$\text{[60,UT]}$, position=below]{G}
\Edge(A)(B)
\Edge(A)(C)
\Edge(C)(B)
\Edge (C)(D)
\Edge(B)(D)
\Edge(E)(D)
\Edge(E)(F)
\Edge(E)(G)
\Edge(G)(F)
\Edge(F)(H)
\end{tikzpicture}
\end{subfigure}
\begin{subfigure}{.24\textwidth} 
\begin{tikzpicture}

\Vertex[size=.1,y=2.2,Pseudo=True,label=Node attr.: $\text{[Age, State]}$, position=right]{0}

\Vertex[size=.3,x=-.2,color=blue,label=$\text{[25,CA]}$, position=above]{A}
\Vertex[size=.3,x=.5,y=1, color=blue, label=$\text{[30,CA]}$, position=above]{B}
\Vertex[size=.3,x=.5,y=-1, color=blue,label=$\text{[35,CA]}$, position=below]{C}
\Vertex[size=.3,x=1.2, color=blue,label=$\text{[40,NV]}$, position=below]{D}
\Vertex[size=.3,x=2.2, color=red,label=$\text{[45,NV]}$, position=below]{E}
\Vertex[size=.3,x=2,y=1, color=red,label=$\text{[35,UT]}$, position=above]{F}
\Vertex[size=.3,x=3,y=1, color=red,label=$\text{[20,ID]}$, position=above]{H}
\Vertex[size=.3,x=3,y=-1, color=red,label=$\text{[60,UT]}$, position=below]{G}
\Edge(A)(B)
\Edge(A)(C)
\Edge(C)(B)
\Edge (C)(D)
\Edge(B)(D)
\Edge(E)(D)
\Edge(E)(F)
\Edge(E)(G)
\Edge(G)(F)
\Edge(F)(H)

\Edge[color=red,style={dashed}, label=$\tilde{e}_1$, position = above](A)(F)
\Edge[color=red,style={dashed}, label=$\tilde{e}_2$, position=right](G)(H)
\end{tikzpicture}
\end{subfigure}
\caption{Original (left) and perturbed (right) friendship network of voters. Each voter is associated with attributes (age, state); and, same color voters belong to the same community (Democrat, Republic). Anomalous edges may appear between nodes of different communities (anomalous edge $\tilde{e}_1$) or between nodes with uncorrelated attributes (anomalous edge $\tilde{e}_2$).}
\label{fig:network}
\end{figure}
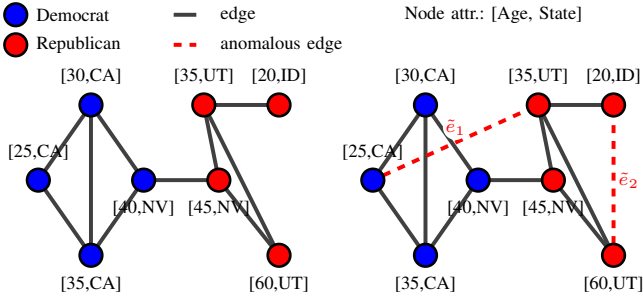

\noindent \textbf{Goal.} Given the perturbed Laplacian matrix $ \tilde{\mathbf{L}} \in \mathbb{R}^{N \times N}$ and the feature matrix $\mathbf{X}$,
this paper aims to unveil the anomalous edges in  $\tilde{\mathcal{E}}$  and recover the unperturbed Laplacian matrix $\mathbf{L}$. Fig. \ref{fig:network} shows a perturbed social voting network.

\subsection{Anomalous edges across communities}
Anomalous edges often connect nodes belonging to different communities \cite{Akoglu2014Graph-basedSurvey, Baingana2016JointNetworks, Chakrabarti2004AutoPart:Detection, Tong2011Non-negativeDetection}. For example, in a typical social voting network, each node is either Republican or Democrat, and connections between nodes across the isle are rare \cite{Akoglu2014Graph-basedSurvey}.  

Factorizing the Laplacian matrix of a graph is known to reveal the community structure  \cite{Baingana2016JointNetworks,  Kuang2012SymmetricClustering,Tong2011Non-negativeDetection, WangStructuralEmbedding}. Specifically, the low-rank approximation of the Laplacian yields the communities, while the low rank captures the number of communities. To retrieve anomalous edges connecting different clusters, it is possible to augment the decomposition with a sparse matrix containing edges that connect different communities, and hence do not obey the low-rank model
\cite{Ioannidis2018CoupledDetection, Baingana2016JointNetworks, Tong2011Non-negativeDetection}. Upon accounting for the low-rank approximation matrix $\mathbf{R}$ and the sparse matrix $\mathbf{S}$, our formulation amounts to
\begin{align}
\label{eq:lowrank}
\min_{\mathbf{S},\mathbf{R}} \; \|\tilde{\mathbf{L}}-\mathbf{S}-\mathbf{R}\|^2_F + \lambda \|\mathbf{S}\|_1 + \mu \|\mathbf{R}\|_* 
\end{align}
where $\| \cdot \|_F$ denotes the Frobenius norm;  $\| \cdot  \|_1$ the $\ell_1$ norm; $\| \cdot \|_*$ the nuclear norm; and, $\lambda,\mu>0$ are properly chosen regularization parameters. The $\ell_1$ norm controls sparsity of $\mathbf{S}$, while the nuclear norm equals the $\ell_1$ norm applied to the vector of singular values of $\mathbf{R}$, which promotes $\mathbf{R}$ to be low rank~\cite{Recht2008NecessaryMinimization}. The nonzero entries of $\mathbf{S}$ reveal the anomalous edges that do not adhere to the low rank model.

Albeit interesting, the approach in \eqref{eq:lowrank} does not leverage the feature matrix $\mathbf{X}$. In the social voting network paradigm, two individuals with opposing political views may be friends because they live in the same state or belong to the same age group (cf. Fig. \ref{fig:network}). The next section outlines our modeling approach that incorporates nodal features when available.

\subsection{Smoothness over features}

Smoothness of signal models has been widely adopted by several learning methods in the form of  regularization for e.g., (semi-)supervised learning over graphs \cite{Dong2016LearningRepresentations}. Intuitively, $\mathbf{X}$ is deemed smooth if features of connected vertices have similar values. This principle suggests, for instance, estimating one person’s age by looking at their friends’ age \cite{Romero2017Kernel-BasedGraphs, Smola2003KernelsGraphs}.

Smooth features among neighboring nodes are captured by the so-termed Laplacian regularizer  \cite{Romero2017Kernel-BasedGraphs, Smola2003KernelsGraphs,ioannidis2018kernellearn}
\begin{equation}
    \frac{1}{2}\sum_{i =1}^n \sum_{j =1}^n a_{ij}\|\mathbf{x}_{i} - \mathbf{x}_{j}\|^2_2 = \text{Tr}(\mathbf{X}^T\mathbf{L}\mathbf{X})
    \label{eq:smooth}
\end{equation}
where $a_{ij} \neq 0$ only if nodes $i$ and $j$ are connected, ${\rm Tr}$ denotes matrix trace, and ${}^T$ stands for transposition. The value of \eqref{eq:smooth} is small if neighboring  vertices have  similar signal values and high otherwise. Thus, minimizing \eqref{eq:smooth} penalizes differences between features of connected nodes. 

\section{Leveraging sparsity, smoothness and low rank}
Incorporating feature smoothness, \eqref{eq:lowrank} boils down to (cf. \eqref{eq:smooth})
\begin{equation}
    \min_{\mathbf{S},\mathbf{R}} \; \|\tilde{\mathbf{L}}-\mathbf{S}-\mathbf{R}\|^2_F + \lambda \|\mathbf{S}\|_1 + \mu \|\mathbf{R}\|_\ast  +\gamma \text{Tr}(\mathbf{X}^{T}\mathbf{R}\mathbf{X})
\label{eq:base2}
\end{equation}
where $\gamma>0$ tunes the degree of smoothness. The non-zero entries of $\mathbf{S}$ reveal edges that do not adhere to the low rank model with smooth features. The next subsections provide alternative formulations of~\eqref{eq:base2}, and corresponding optimization solvers with complementary benefits.

\subsection{Anomalous edge identification via low rank factorization}
As both the nuclear and $\ell_1$ norms are non-differentiable at the origin, the optimization problem (\ref{eq:base2}) is convex but non-smooth. Although iterative solvers of \eqref{eq:base2} are available, they rely on performing a costly singular value decomposition per iteration, which incurs prohibitive computational complexity, especially for large-scale graphs \cite{Mardani2013DynamicRank, Mardani2013RecoveryAnomalies}. 

A prudent strategy to bypass this hurdle relies on the factorization $\mathbf{R} = \mathbf{U}\mathbf{V}^T$, where $\mathbf{U},\mathbf{V}$ are $N \times R$ matrices with $R$ denoting an upper bound on the rank  of $\mathbf{R}$. Specifically, it holds that \cite{Recht2010GuaranteedMinimization, Re2011ParallelCompletion}
\begin{equation}
\|\mathbf{R}\|_* = \min_{\mathbf{U},\mathbf{V}} \;  \|\mathbf{U}\|^2_F + \|\mathbf{V}\|^2_F~~~~ \text{s.t.} \; \mathbf{R}=\mathbf{U}\mathbf{V}^T \;.
\label{eq:lowr}
\end{equation}
Combining~\eqref{eq:lowr} with~\eqref{eq:base2}, the optimization in \eqref{eq:base2} reduces to
\begin{equation}
\begin{split}
    \min_{\mathbf{S},\mathbf{U},\mathbf{V}} \; & f(\mathbf{U},\mathbf{V},\mathbf{S})  :=  \\ 
    & \|\tilde{\mathbf{L}}-\mathbf{S}-\mathbf{U}\mathbf{V}^T\|^2_F + \lambda \|\mathbf{S}\|_1 + \mu (\|\mathbf{U}\|^2_F + \|\mathbf{V}\|^2_F) \\
    & +\gamma \text{Tr}(\mathbf{X}^{T}(\mathbf{U}\mathbf{V}^{T})\mathbf{X}) \;.
\end{split}
\label{probform}
\end{equation}
Although (\ref{probform}) is non-convex, alternating minimization solvers can be employed with 
closed-form updates. Different from \eqref{eq:base2}, problem~\eqref{probform} can be solved in a distributed fashion.  We developed an alternating least-squares (ALS) solver of (\ref{probform}) whose steps are listed under Algorithm \ref{alg:als}. Our ALS is provably convergent, but the proof is omitted due to space limitations.


\subsection{Anomalous edge identification aided by Laplacian recovery}



The formulation in (\ref{eq:base2}), focused solely on retrieving the anomalous edges. However, one can also approximate the  \emph{unperturbed} graph structure, and thus aid the identification of anomalous edges. This subsection deals with robust estimation of the nominal graph topology too, and further improving identification of the anomalous edges. 

Graph connections capture correlations among nodal feature vectors; see~\cite{Giannakis2018TopologyDynamics} for a comprehensive survey. Learning the graph structure by leveraging the statistical properties of the features will endow estimation with increased robustness to anomalies. Popular approaches to estimating graph structure are closely related to the sparse inverse covariance matrix estimation \cite{Lake10discoveringstructure,Yuan2007modelselection, Benerjee2008ModelSelection}. The graph Laplacian can be approximated similarly by estimating the sparse inverse covariance matrix \cite{fht2008biostats,Lake10discoveringstructure,Zhao2019OptimizationMm}. Specifically, the sparse inverse covariance selection problem \cite{fht2008biostats} is formulated  as follows
\begin{equation}
    \min_{\mathbf{\Theta}\succeq \mathbf{0}} \;  \text{Tr}(\mathbf{X}\mathbf{X}^T\mathbf{\Theta}) - \log \det (\mathbf{\Theta}) + \beta \|\mathbf{\Theta}\|_1 
\label{eq:covarian}
\end{equation}
where $\text{Tr}(\mathbf{X}\mathbf{X}^T\mathbf{\Theta}) - \log \det (\mathbf{\Theta})$ minimizes the negative log-likelihood, while $\|\mathbf{\Theta}\|_1$ promotes sparsity. 
The term $\text{Tr}(\mathbf{X}\mathbf{X}^T\mathbf{\Theta})$ is closely related to the smoothing term $\text{Tr}(\mathbf{X}^T\mathbf{R}\mathbf{X})$ of \eqref{eq:base2}, and holds the feature correlation information. Although $\mathbf{\Theta}$ captures feature correlations ($\mathbf{\Theta}_{ij} \neq 0 $ if nodes $i$ and $j$ have conditionally correlated features), it is not necessarily a valid Laplacian.

To cope with this challenge, we couple~\eqref{eq:lowrank} and~\eqref{eq:covarian}, and introduce due constraints to recover the unperturbed graph Laplacian and the anomalies
 \begin{subequations}
\begin{align}
    \min_{\mathbf{R},\mathbf{S}, \mathbf{\Theta} \succeq \mathbf{0}} & \;   \|\tilde{\mathbf{L}}-\mathbf{S}-\mathbf{R}\|^2_F + \lambda \|\mathbf{S}\|_1 + \mu \|\mathbf{R}\|_* \notag \\
    \; & + \alpha ( \text{Tr}(\mathbf{XX}^T\mathbf{\Theta}) - \log \det (\mathbf{\Theta}) + \beta \|\mathbf{\Theta}\|_1 ) \notag \\
    \; & + \kappa \|\mathbf{R}-\mathbf{\Theta}\|_F^2 \\
    \text{subject to} & \notag \\
    & \;\text{Tr}(\mathbf{R}) \geq m \label{prob:Laplc0}\\
     & \;\mathbf{R} \mathbf{1}= \mathbf{0} \label{prob:Laplc1} \\ 
     & \;\mathbf{R}_{ij} \leq 0, \; i\neq j \label{prob:Laplc2}
\end{align}
\label{prob:Lapl-couple}
\end{subequations}
where \eqref{prob:Laplc0}, \eqref{prob:Laplc1}, \eqref{prob:Laplc2} constrain the learned $\mathbf{R}$ to be positive semidefinite and thus a valid Laplacian, while  $\|\mathbf{R}-\mathbf{\Theta}\|_F^2$ seeks a Laplacian that stays close to $\mathbf{\Theta}$ but also couples \eqref{eq:lowrank} with~\eqref{eq:covarian} (closeness is tuned by the scalar $\kappa >0$). The number of  edges in the sought Laplacian is benchmarked by $m$ (cf. \eqref{prob:Laplc0}), while $\alpha>0$ weighs the degree of smoothness in~\eqref{eq:lowrank} and~\eqref{eq:covarian}.

The problem in~\eqref{prob:Lapl-couple} is convex and can be solved using either well known optimization solvers (e.g interior point schemes), or, the alternating direction method of multipliers (ADMM) method that can also afford for distributed implementation; we will henceforth refere to this ADMM solver of \eqref{prob:Lapl-couple}) as Algorithm 2. Note also that upon  recovering the Laplacian, the adjacency matrix can be readily obtained. 

\begin{algorithm}[h]
\caption{\strut ALS for (\ref{probform})}\label{alg:als}
\begin{algorithmic}[1]
\small
\Initialize{$l=0, \mathbf{U}^{(l)}, \mathbf{V}^{(l)}, \mathbf{S}^{(l)}$}
\While{not converged}
\State $\mathbf{U}^{(l+1)} = \argminl_{\mathbf{U}} f(\mathbf{U},\mathbf{V}^{(l)},\mathbf{S}^{(l)}) $
\State $\mathbf{V}^{(l+1)} = \argminl_{\mathbf{V}} f(\mathbf{U}^{(l+1)},\mathbf{V},\mathbf{S}^{(l)}) $
\State $\mathbf{S}^{(l+1)} = \argminl_{\mathbf{S}} f(\mathbf{U}^{(l+1)},\mathbf{V}^{(l+1)},\mathbf{S}) $ 
\State $l=l+1$
\EndWhile
\vspace{-0.1cm}
\end{algorithmic}
\end{algorithm}

\noindent\textbf{Computational complexity}. 
The $\mathbf{U}$ and $\mathbf{V}$ updates under Algorithm \ref{alg:als} are available in closed form, while $\mathbf{S}$ is updated entry wise using the soft threshold operator \cite{Parikh2014ProximalAlgorithms}. Considering $R \ll F \ll N$, the most computationally expensive operation is matrix multiplication between $N \times N$ and $N \times R$ matrices. Therefore, the overall complexity per step is $\mathcal{O}(N^2R)$~\cite{Davie2013ImprovedMultiplication}.

Algorithm 2 minimizes a convex function under \eqref{prob:Laplc0}, \eqref{prob:Laplc1}, and \eqref{prob:Laplc2} that can be expressed as linear constraints \cite{Dong2014LearningRepresentations}. This implies that \eqref{prob:Lapl-couple} can be efficiently solved in $\mathcal{O}(\log (\frac{1}{\epsilon}))$ interior point sub-iterations; see e.g., 
\cite{Boyd2004ConvexOptimization,Diamond2016CVXPY:Optimization},  \cite{Ragnavendra2008OptimalCSP,Burer2003AFactorization, Klerk2002AspectsApplications}. Accounting also for the SVD computations $\mathcal{O}(N^3)$, and matrix multiplications, the overall complexity is $\mathcal{O}(\log (\frac{1}{\epsilon}) N^3)$.

\section{Numerical Tests} \label{sec:Exper}
This section evaluates the anomaly identification performance of the novel methods with real and synthetic graph datasets. 

\textbf{Stochastic block model (SBM)}. 
SBMs are synthetic networks that consist of $C$ communities. The feature matrix is formed by the $F$ eigenvectors of the Laplacian matrix that correspond to the 
$F$ smallest eigenvalues. Such features are smooth with respect to the graph \cite{Romero2017Kernel-BasedGraphs}.
The generated stochastic block model networks are denoted as SBM($C$,$F$).

\textbf{Aarhus network}. The Aarhus dataset is a social network that consists of 32 employees (nodes) from the Computer Science Department of Aarhus University and represents their lunch relations (edges)~\cite{BayramMaskInference}. The $32 \times 26$ feature matrix $\mathbf{X}$ is a binary matrix that represents the lunch relationships between the 32 employees from Computer Science department and another group of 26 people. 

\subsection{Evaluation results}
\begin{table*}[ht]
\centering
\caption{Best averaged metrics achieved on each graph configuration for each model. Bold font indicates the highest value of each metric in a given network. } \label{table:Perf1}
\scalebox{1.2}{
\begin{tabular}{|c|| r| r| r| r | r | r |  r | r |} 
\hline
\multirow{2}{*}{Dataset} & \multicolumn{2}{c|}{Random guess} & \multicolumn{2}{c|}{Baseline} & \multicolumn{2}{c|}{Algorithm 1} & \multicolumn{2}{c|}{ Algorithm 2 }  \\ \cline{2-9}
  & H@10& MRR & H@10& MRR & H@10 & MRR &  H@10 & MRR  \\ 
\hline 
\hline
 SBM(4,4) & 6.6 & 0.7 & 24.0  & 34.8 & \textbf{57.0} & 58.6 & 55.9  &  \textbf{72.6} \\
\hline 
SBM(6,3)  & 4.0 & 0.4 &14.0 & 22.1  & \textbf{46.7} & 37.6 & 42.6 & \textbf{64.6}  \\ 
\hline 
 SBM(8,4) & 0.3 & $\sim$ 0.0 &11.0 & 20.5 &  24.0 & 24.2 & \textbf{42.0} & \textbf{55.5} \\
\hline 
Aarhus & 13.7 & 1.4 & 38.0 & 25.6 & \textbf{72.0}  & 46.9 & 66.0 & \textbf{88.4}  \\
\hline
 \end{tabular}
}
\end{table*}

\begin{figure*}[t]
\centering
\begin{subfigure}{.24\textwidth}
  \centering
  \includegraphics[width=1.1\textwidth]{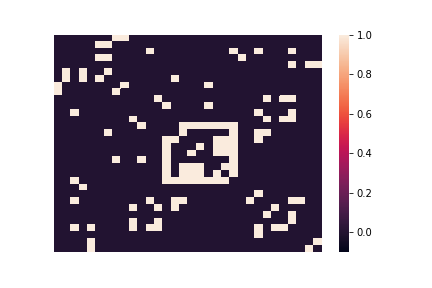}  
\end{subfigure}
\begin{subfigure}{.24\textwidth}
  \centering
  \includegraphics[width=1.1\textwidth]{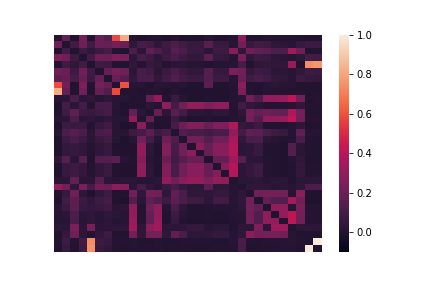}  
\end{subfigure}
\begin{subfigure}{.24\textwidth}
  \centering
  \includegraphics[width=1.1\textwidth]{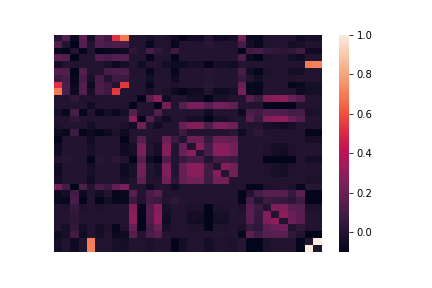}  
\end{subfigure}

\begin{subfigure}{.24\textwidth}
  \centering
  \includegraphics[width=1.1\textwidth]{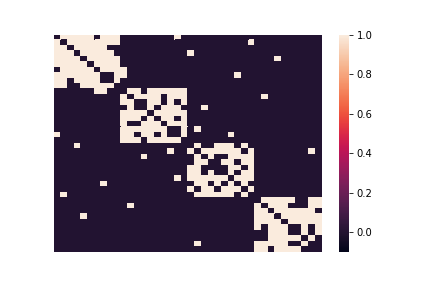}  
  
\end{subfigure}
\begin{subfigure}{.24\textwidth}
  \centering
  \includegraphics[width=1.1\textwidth]{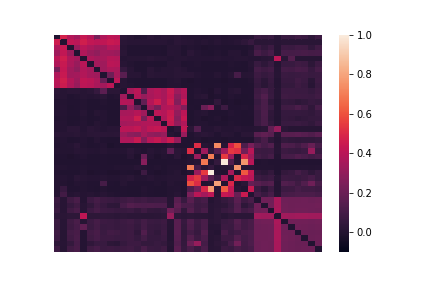}  
  
\end{subfigure}
\begin{subfigure}{.24\textwidth}
  \centering
  \includegraphics[width=1.1\textwidth]{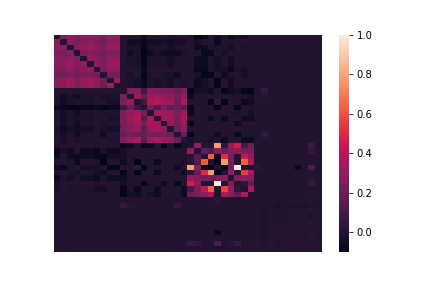}  
  
\end{subfigure}
\caption{Aarhus (top) and SBM($4,4$) (bottom) networks heatmap of the original (left) adjacency. The estimated one (middle) from Method 2 is closer to the original than the baseline estimation (right).}
\label{fig:Lapl}
\end{figure*}
In the SBM, anomalous edges are randomly added between nodes that belong to different communities, while in the Aarhus dataset, anomalous edges are inserted randomly between any two nodes. After the optimization, edges with negative (nonzero) values in $\mathbf{S}$ are collected in the set $\hat{\mathcal{E}}$. Edges in $\hat{\mathcal{E}}$ are sorted in ascending order so that possible anomalous edges are ordered by probability of correct detection. The tests are repeated 10 times and the average percentage values of the following metrics are reported.

\textbf{Hit rate@10} (H@10), which indicates how many edges are correctly identified as anomalous, given the first 10 candidate anomalous edges of $\hat{\mathcal{E}}$. 

\textbf{Mean reciprocal rank} (MRR). Each edge of $\hat{\mathcal{E}}$ receives a reciprocal rank score (if correctly identified as anomalous) based on its position in $\hat{\mathcal{E}}$: 1 for first place, 1/2 for second place, 1/3 for third place, and so on. MRR is the average of reciprocal ranks of all correctly identified anomalous edges. 


Table \ref{table:Perf1}  illustrates the \textit{best} evaluation results for different methods after systematic parameter search.  
The following  baseline methods are considered.  The method in~\eqref{eq:lowrank} that is employed by various approaches, e.g \cite{Baingana2016JointNetworks},  and does not consider feature smoothness and a random-guess method that selects a random subset of the edges as anomalous.
 
 Both proposed methods significantly outperform the baseline, which speaks for the importance of smoothness constraints for anomaly identification. The improvements on H@10 accuracy metric compared to the baseline ranges from 1.9 times higher (Aarhus,  Algorithm 1) to 3.8 times higher (SBM(8,4),  Algorithm 2). Improvements on MRR ranges from 1.2 times (SBM(8,4),  Algorithm 1) to 3.5 times higher (Aarhus,  Algorithm 2). The highest values of H@10 and MRR are observed in the Aarhus dataset ($\sim$ 72\% and 88.4\% respectively). In all cases, our second algorithm outperforms the first one in the MRR metric due to its ability to approximate the exact topology of the original graph, which enhances the anomaly identification performance. On the other hand, although Algorithm 2 achieves better anomaly identification performance, it is less computationally efficient compared to Algorithm 1.

Finally, Figure \ref{fig:Lapl} illustrates the estimated graph topology by the second method, compared to the ground truth topology on two datasets. As a baseline method, method \eqref{eq:covarian} is used, which is employed by various approaches, e.g. 
\cite{Dong2016LearningLaplac}. These qualitative results confirm that the second method captures the main topological structure of the original graph, while its topology estimation is closer to the ground truth topology than the baseline method.
\vspace{-0.1cm}
\section{Conclusions and future work}\vspace{-0.1cm}
The present work aims at identifying anomalous edges on attributed networks. Two novel approaches are formulated to jointly leverage the low rank plus sparse decomposition of the Laplacian of the perturbed graph, along with feature smoothness. 
Extensive numerical tests corroborate that the novel adaptation of feature smoothness markedly improves the anomaly identification performance and the proposed methods outperform the baseline. 
Future research directions include robust topology identification, robust community detection by employing ego-tensors~\cite{sheikholeslami2017overlapping}, and anomaly identification on dynamic and knowledge graphs.

\bibliographystyle{ieeetr}
\bibliography{Anomaly}\noindent 

\end{document}

%% file: include.tex
\usepackage{fixltx2e}

\usepackage{graphicx}

\usepackage{subcaption}              

\usepackage[utf8]{inputenc}

\usepackage{amsfonts}
\usepackage{amsmath}
\usepackage{mathtools}

\usepackage{amssymb}

\usepackage{bm}

\usepackage{color,verbatim}
\usepackage{multirow}
\usepackage{accents}

\usepackage{theoremref}

\usepackage{flushend}

\usepackage{url}

\newcounter{rulecounter}
\newcommand{\resetrule}{ \setcounter{rulecounter}{0}}
\resetrule

\newsavebox{\selvestebox}
\newenvironment{colbox}[1]
  {\newcommand\colboxcolor{#1}%
   \begin{lrbox}{\selvestebox}%
   \begin{minipage}{\dimexpr\columnwidth-2\fboxsep\relax}}
  {\end{minipage}\end{lrbox}%
   \begin{center}
   \colorbox{\colboxcolor}{\usebox{\selvestebox}}
   \end{center}}

\definecolor{orange}{rgb}{1,0.8,0}
\definecolor{gray}{rgb}{.9,0.9,0.9}
\definecolor{darkgray}{rgb}{.3,0.3,0.3}
\definecolor{darkblue}{rgb}{.1,0.0,0.3}
\definecolor{lightblue}{rgb}{0.7,0.7,1}
\definecolor{lightred}{rgb}{1,0.7,.7}
\definecolor{purple}{RGB}{204,153,255}
\definecolor{lightgray}{rgb}{.95,0.95,0.95}
\definecolor{lightgreen}{rgb}{0.3,0.5,0.3}
\definecolor{darkgreen}{rgb}{0.05,0.3,0.05}

\newtheorem{myproposition}{Proposition}
\newtheorem{myremark}{Remark}
\newtheorem{myproblemstatement}{Problem Statement}
\newtheorem{mylemma}{Lemma}
\newtheorem{mytheorem}{Theorem}
\newtheorem{mydefinition}{Definition}
\newtheorem{mycorollary}{Corollary}

\usepackage{pgfplots}
\pgfplotsset{compat=newest}
\usetikzlibrary{plotmarks}
\usetikzlibrary{positioning}
\usetikzlibrary{arrows.meta}
\usepgfplotslibrary{patchplots}
\usepackage{grffile}

\pgfplotsset{plot coordinates/math parser=false}
\newlength\mywidth
\newlength\myheight
\newlength\mywidths
\newlength\myheights
\newlength\mywidthss
\newlength\myheightss
\definecolor{mycolor1}{rgb}{0.00000,0.44700,0.74100}%
\definecolor{mycolor2}{rgb}{0.85000,0.32500,0.9800}%
\definecolor{mycolor3}{rgb}{0.92900,0.69400,0.12500}%
\definecolor{mycolor4}{rgb}{0.89400,0.18400,0.15600}%
\definecolor{mycolor5}{rgb}{0.46600,0.67400,0.18800}%
\definecolor{mycolor6}{rgb}{0.30100,0.74500,0.93300}%
\definecolor{mycolor7}{rgb}{0.63500,0.07800,0.18400}%

\definecolor{AGNN}{rgb}{1,0.04700,0.04100}%
\definecolor{GCN}{rgb}{0.05000,0.32500,0.89800}%
\definecolor{AGNNnf}{rgb}{1,0.8,0.119800}%
\definecolor{Mune}{rgb}{0.2,0.800,1}%

\definecolor{TightHann}{rgb}{0.92900,0.69400,0.12500}%
\definecolor{pDiffScater}{rgb}{0.49400,0.18400,0.55600}%
\definecolor{pMonicCubic}{rgb}{0.46600,0.67400,0.18800}%
\definecolor{pTightHann}{rgb}{0.30100,0.74500,0.93300}%

\pgfplotscreateplotcyclelist{colorlist}{%
color=AGNN,line width=2.0pt,
solid, every mark/.append style={solid, fill=AGNN}, mark=*\\%
color=GCN,line width=2.0pt,dotted, every mark/.append style={solid, fill=GCN}, mark=square*\\%
color=AGNNnf,line width=2.0pt,densely dotted, every mark/.append style={solid, fill=AGNNnf}, mark=otimes*\\%
color=Mune,line width=2.0pt,loosely dotted, every mark/.append style={solid, fill=Mune}, mark=triangle*\\%
color=mycolor5,line width=2.0pt,
dashed, every mark/.append style={solid, fill=mycolor5},mark=diamond*\\%
color=mycolor6,line width=2.0pt,loosely dashed, every mark/.append style={solid, fill=mycolor6},mark=*\\%
color=mycolor7,densely dashed, every mark/.append style={solid, fill=mycolor7},line width=2.0pt,mark=square*\\%
dashdotted, every mark/.append style={solid, fill=mycolor7},mark=otimes*\\%
dashdotdotted, every mark/.append style={solid},mark=star\\%
densely dashdotted,every mark/.append style={solid, fill=gray},mark=diamond*\\%
}
\pgfplotscreateplotcyclelist{my black white}{%
solid, every mark/.append style={solid, fill=gray}, mark=*\\%
dotted, every mark/.append style={solid, fill=gray}, mark=square*\\%
densely dotted, every mark/.append style={solid, fill=gray}, mark=otimes*\\%
loosely dotted, every mark/.append style={solid, fill=gray}, mark=triangle*\\%
dashed, every mark/.append style={solid, fill=gray},mark=diamond*\\%
loosely dashed, every mark/.append style={solid, fill=gray},mark=*\\%
densely dashed, every mark/.append style={solid, fill=gray},mark=square*\\%
dashdotted, every mark/.append style={solid, fill=gray},mark=otimes*\\%
dashdotdotted, every mark/.append style={solid},mark=star\\%
densely dashdotted,every mark/.append style={solid, fill=gray},mark=diamond*\\%
}
\usetikzlibrary{decorations.pathreplacing}
\definecolor{lavander}{cmyk}{0,0.48,0,0}
\definecolor{violet}{cmyk}{0.79,0.88,0,0}
\definecolor{burntorange}{cmyk}{0,0.52,1,0}
\definecolor{mygreen}{rgb}{0,1,0}%
\definecolor{myred}{rgb}{1,0,0}%

\tikzstyle{nnblock}=[draw,rectangle,  rounded corners,text=black,
minimum width=30pt,minimum height=60pt]
\tikzstyle{esl}=[draw,rectangle,  rounded corners,text=black,
minimum width=60pt,minimum height=30pt]
\tikzstyle{grnn}=[draw,rectangle,  rounded corners,text=black,
minimum width=30pt,minimum height=15pt]
\tikzstyle{graphsampler}=[draw,rectangle,  fill=orange,fill opacity=0.1, rounded corners,text=orange,
minimum width=30pt,minimum height=15pt]
\tikzstyle{outblock}=[draw,rectangle, color=red, rounded corners,text=black,
minimum width=10pt,minimum height=60pt]
\tikzstyle{nam}=[rectangle, fill=orange,fill opacity=0.1, rounded corners,text=orange,
minimum width=8pt,minimum height=58pt]
\tikzstyle{gam}=[fill=blue,fill opacity=0.1, rectangle,  rounded corners,text=black,
minimum width=8pt,minimum height=58pt]
\tikzstyle{fam}=[fill=green,fill opacity=0.1, rectangle,  rounded corners,text=black,
minimum width=8pt,minimum height=58pt]

\tikzstyle{peers}=[draw,circle,bottom color=myred,
                  top color= white, text=violet,minimum width=8pt]
\tikzstyle{superpeers}=[draw,circle, left color=mygreen,
                       text=violet,minimum width=8pt]

\tikzstyle{peers3}=[draw,circle,bottom color=myred,
                  top color= white, text=violet,minimum width=12pt]
\tikzstyle{superpeers3}=[draw,circle, left color=mygreen,
                       text=violet,minimum width=12pt]
\tikzstyle{peers2}=[draw,circle,bottom color=myred,
                  top color= white, text=violet,inner sep=0pt,minimum size=0.1cm]
\tikzstyle{superpeers2}=[draw,circle, left color=mygreen,
                       text=violet,inner sep=0pt,minimum size=0.1cm]
                    
\tikzstyle{peers1}=[draw,circle,bottom color=myred,
                  top color= white, text=violet,minimum width=4pt]
\tikzstyle{superpeers1}=[draw,circle, left color=mygreen,
                       text=violet,minimum width=4pt]
\tikzstyle{legendsp}=[rectangle, draw,  rounded corners,
                     thin,bottom color=mygreen, top color=white,
                     text=black, minimum width=2cm]
\tikzstyle{legendp}=[rectangle, draw,  rounded corners, thin,
                     bottom color=myred, top color= white,
                     text= black, minimum width= 2cm]
\tikzstyle{legend_general}=[rectangle, rounded corners, thin,
                           burntorange, fill= white, draw, text=violet,
                           minimum width=2.5cm, minimum height=0.8cm]
\usetikzlibrary{shapes.misc, positioning}

\usetikzlibrary{calc,trees}